\documentclass[aps,showpacs,preprintnumbers,amsmath, amssymb]{revtex4}

\usepackage{float}
\usepackage{graphics,epsfig}
\usepackage{graphicx}
\usepackage{dcolumn}
\usepackage{bm}

\begin{document}
\baselineskip=0.8 cm

\title{{\bf A no-go theorem for scalar fields with couplings from Ginzburg-Landau models}}
\author{Guohua Liu$^{1}$\footnote{liuguohua1234@163.com}}
\author{Yan Peng$^{2}$\footnote{yanpengphy@163.com}}
\affiliation{\\$^{1}$ School of Physics and Physical Engineering, Qufu Normal University, Qufu, Shandong, 273165, China}
\affiliation{\\$^{2}$ School of Mathematical Sciences, Qufu Normal University, Qufu, Shandong 273165, China}

\vspace*{0.2cm}
\begin{abstract}
\baselineskip=0.6 cm
\begin{center}
{\bf Abstract}
\end{center}

Recently Hod proved a no-go theorem that
static scalar fields cannot form
spherically symmetric boson stars
in the asymptotically flat background.
On the other side, scalar fields can be coupled to the gradient
according to next-to-leading order Ginzburg-Landau models.
In the present work, we extend Hod's discussions by considering
couplings between static scalar fields and the field
gradient. For a non-negative coupling parameter,
we show that there is no asymptotically flat spherically symmetric
boson stars made of coupled static scalar fields.

\end{abstract}

\maketitle
\newpage
\vspace*{0.2cm}

\section{Introduction}

There is growing evidence supporting the existence
of fundamental scalar fields in nature. The systems composed
of scalar fields have attracted a lot of attention
from mathematicians and physicists. One of the most famous
properties of such systems is the no hair theorem,
which states that static scalar fields cannot exist
outside asymptotically flat black hole horizons,
see references \cite{Bekenstein}-\cite{Brihaye} and reviews \cite{Bekenstein-1,CAR}.
Lately, it was found that no scalar field property
also appears in curved horizonless spacetimes \cite{Hod-6}-\cite{LR1}.

A well known horizonless configuration is the boson star made of
scalar fields. Theoretically, the scalar fields can be stationary or static.
It was found that boson stars
can be constructed with stationary scalar fields \cite{FE,DAER}.
Then it is interesting to examine whether
static scalar fields can form boson stars.
In the flat spacetime, boson stars cannot be made of
static scalar fields due to Derrick's theorem \cite{GHD}.
In the asymptotically flat background, Hod proved that
boson stars cannot be composed of static scalar fields \cite{ng1,ng2}.
In the asymptotically AdS background,
this no boson star property also appears \cite{ng3}.
On the other side, in next-to-leading order Ginzburg-Landau
family of models for a BCS superconductor, there is
a new term with scalar fields coupled to the gradient \cite{cond1,cond2,cond3}
and the couplings play an important role in scalar condensations \cite{GLYP}.
Along this line, it is interesting to study the no
boson star property with coupled static scalar fields.

In the following, we plan to extend the discussion
in \cite{ng1} by considering static scalar fields
coupled to the gradient in the asymptotically flat background.
We show that boson stars cannot be made of static scalar fields
with a non-negative coupling parameter.
We summarize main results in the last section.

\section{No static boson star for coupled scalar fields}

In the extended Ginzburg-Landau models, a new term with scalar fields
coupled to the gradient appears \cite{cond1,cond2,cond3}.
In this work, we study the system with coupled static scalar fields
in the curved spacetime. The Lagrangian density describing
scalar fields coupled to the gradient is \cite{dg,sh,Rogatko1}
\begin{eqnarray}\label{lagrange-1}
\mathcal{L}=R-|\nabla_{\alpha} \psi|^{2}-\xi\psi^{2}|\nabla_{\alpha} \psi|^{2}-V(\psi^{2}),
\end{eqnarray}
where R corresponds to the Ricci curvature and $\xi$ is the coupling parameter.
In order to obtain results in this work, we assume that the coupling parameter
is non-negative as $\xi>0$. We study radial direction depending scalar
fields expressed as $\psi=\psi(r)$.

The potential $V(\psi^2)$ is positive semidefinite
and increases as a function of $\psi^2$ satisfying
\begin{eqnarray}\label{lagrange-1}
V(0)=0~~~~and~~~~\dot{V}=\frac{dV(\psi^2)}{d(\psi^2)}> 0.
\end{eqnarray}
The free scalar fields with $V(\psi^2)=\mu^2\psi^2$
apparently satisfy the relation (2) as
$V(0)=0$ and $\dot{V}=\mu^2>0$. Here $\mu$
is the nonzero scalar field mass.

The metric of spherically symmetric boson star can be expressed as
 \cite{mr1,mr2,fc,Basu,Rogatko,Peng Wang}
\begin{eqnarray}\label{AdSBH}
ds^{2}&=&-ge^{-\chi}dt^{2}+\frac{dr^{2}}{g}+r^{2}(d\theta^2+sin^{2}\theta d\phi^{2}).
\end{eqnarray}
$\chi$ and $g$ are functions of the coordinate r.
$\theta$ and $\phi$ are angular coordinates.

The Lagrangian density (1) yields the scalar field equation
\begin{eqnarray}\label{BHg}
(1+\xi\psi^2)\psi''+[(1+\xi\psi^2)(\frac{2}{r}-\frac{\chi'}{2}+\frac{g'}{g})+2\xi \psi \psi']\psi'-(\xi \psi'^{2}+\frac{\dot{V}}{g})\psi=0.
\end{eqnarray}

At spatial infinity, metric functions asymptotically behave as \cite{ng1}
\begin{eqnarray}\label{AdSBH}
\chi\rightarrow 0,~~~~~~g\rightarrow 1~~~~~~for~~~~~~r\rightarrow \infty.
\end{eqnarray}

And near the origin, metric functions are characterized by \cite{ng1}
\begin{eqnarray}\label{AdSBH}
\chi'\rightarrow 0,~~~~~~g\rightarrow 1+O(r^2)~~~~~~for~~~~~~r\rightarrow 0.
\end{eqnarray}

The corresponding energy density is
\begin{eqnarray}\label{BHg}
\rho=-T^{t}_{t}=g\psi'^{2}+\xi g\psi^{2}\psi'^{2}+V(\psi^{2}).
\end{eqnarray}

The finite gravitational mass condition $M=\int_{0}^{\infty}4\pi r^{2}\rho dr< \infty$
implies that $r^{2}\rho$ decreases faster than $\frac{1}{r}$ around the infinity.
Then we arrive at the asymptotical behavior
\begin{eqnarray}\label{BHg}
r^{3}\rho\rightarrow 0~~~~~for~~~~~r\rightarrow \infty.
\end{eqnarray}

According to relations (2), (7) and (8), we obtain the
infinity vanishing condition
\begin{eqnarray}\label{InfBH}
&&\psi(\infty)=0.
\end{eqnarray}

The scalar field equation near the origin is
\begin{eqnarray}\label{BHg}
(1+\xi\psi^2)\psi''+\frac{2}{r}(1+\xi\psi^2)\psi'-(\xi \psi'^{2}+\dot{V})\psi=0.
\end{eqnarray}

Around the origin, we set the scalar field in the general form $\psi(r)=r^{s}(a_{0}+a_{1}r+a_{2}r^2+\ldots)$ \cite{mr5}.
Putting this expression into equation (10) and considering the leading order, we find
the relation $s=0$ and the solution satisfies the near origin asymptotical behavior
\begin{eqnarray}\label{BHg}
\psi(r)=a[1+\frac{\dot{V}(a^2)}{6(1+\xi a^2)}\cdot r^2]+O(r^3),
\end{eqnarray}
where $a$ is the scalar field value $\psi(0)$ at the origin.

In the case of $a=\psi(0)=0$, also considering the relation (9),
we deduce that the eigenfunction $\psi(r)$ must possess at least
one extremum point $r_{peak}$ \cite{Hod-6}.
The scalar fields are characterized by
\begin{eqnarray}\label{InfBH}
\{ \psi^2>0,~~~~\psi\psi'=0~~~~and~~~~\psi\psi''\leqslant0\}~~~~for~~~~r=r_{peak}.
\end{eqnarray}
At $r=r_{peak}$, the relation (12) yields an inequality
\begin{eqnarray}\label{BHg}
(1+\xi\psi^2)\psi\psi''+[(1+\xi\psi^2)(\frac{2}{r}-\frac{\chi'}{2}+\frac{g'}{g})+2\xi \psi \psi']\psi\psi'-(\xi \psi'^{2}+\frac{\dot{V}}{g})\psi^2<0.
\end{eqnarray}

In the case of $a>0$, (11) yields $\psi'(0)=0$ and $\psi''(0)=\frac{a\dot{V}(a^2)}{3(1+\xi a^2)}>0$,
which implies $\psi' > 0$ for $r>0$ in the near origin region.
The scalar field eigenfunction $\psi(r)$ increases to be more positive around
the origin and finally approaches zero at the infinity.
So one deduces that the scalar field $\psi(r)$ reaches
a positive local maximum value at one extremum point $\tilde{r}_{peak}$.
At this point, the eigenfunction is characterized by
 \begin{eqnarray}\label{InfBH}
\{ \psi>0,~~~~\psi'=0~~~~and~~~~\psi''\leqslant0\}~~~~for~~~~r=\tilde{r}_{peak}.
\end{eqnarray}
At $r=\tilde{r}_{peak}$, the relation (14) yields the inequality
\begin{eqnarray}\label{BHg}
(1+\xi\psi^2)\psi''+[(1+\xi\psi^2)(\frac{2}{r}-\frac{\chi'}{2}+\frac{g'}{g})+2\xi \psi \psi']\psi'-(\xi \psi'^{2}+\frac{\dot{V}}{g})\psi<0.
\end{eqnarray}

And in another cases with $a<0$, one can deduce the
existence of one extremum point $\bar{r}_{peak}$,
where the scalar field $\psi(r)$ reaches
a negative local minimum value.
At this point, the relation is
\begin{eqnarray}\label{InfBH}
\{ \psi<0,~~~~\psi'=0~~~~and~~~~\psi''\geqslant0\}~~~~for~~~~r=\bar{r}_{peak}.
\end{eqnarray}
At $r=\bar{r}_{peak}$, the characteristic inequality is
\begin{eqnarray}\label{BHg}
(1+\xi\psi^2)\psi''+[(1+\xi\psi^2)(\frac{2}{r}-\frac{\chi'}{2}+\frac{g'}{g})+2\xi \psi \psi']\psi'-(\xi \psi'^{2}+\frac{\dot{V}}{g})\psi>0.
\end{eqnarray}

Relations (13), (15) and (17) are in contradiction with the scalar field equation (4).
So the scalar field equation cannot be respected at the corresponding extremum
points. We therefore conclude that
spherically symmetric asymptotically flat boson stars cannot
be constructed with static coupled scalar fields with
non-negative coupling parameters.

\section{Conclusions}

We studied the nonexistence of boson stars composed
of static scalar fields in the spherically symmetric
asymptotically flat background. We considered static
scalar fields coupled to the field gradient,
where the coupling term also appears in the
next-to-leading order Ginzburg-Landau models. For non-negative
coupling parameters, we obtained the characteristic
inequalities (13), (15) and (17) at corresponding extremum points.
However, these relations are in contradiction
with the static scalar field equation (4).
As a summary, we proved the nonexistence of
spherically symmetric asymptotically flat
static boson stars made of scalar fields with
a non-negative coupling parameter.

\begin{acknowledgments}

This work was supported by a grant from Qufu Normal University
of China under Grant No. xkjjc201906. This work was also supported by the Youth Innovations and Talents Project of Shandong
Provincial Colleges and Universities (Grant no. 201909118).

\end{acknowledgments}

\end{document}